\documentstyle[preprint,tighten,aps]{revtex}
\begin{document}
\preprint{TRI-PP-96-28, MKPH-T-96-18}
\draft
\title{Virtual Compton scattering off spin-zero particles at low energies}
\author{H.\ W.\ Fearing}
\address{TRIUMF, 4004 Wesbrook Mall, Vancouver, British Columbia,
Canada V6T 2A3}
\author{S.\ Scherer}
\address{Institut f\"ur Kernphysik, Johannes Gutenberg-Universit\"at,
J.\ J.\ Becher-Weg 45, D-55099 Mainz, Germany}
\date{July 26, 1996}
\maketitle

\begin{abstract}
   We discuss the low-energy behavior of the virtual Compton scattering
amplitude off a spin-zero target.
   We first compare various methods of obtaining a low-energy expression
based either on the soft-photon approximation or the use of Ward-Takahashi
identities.
    We point out that structure-dependent terms are defined
with respect to a low-energy approximation of the pole terms which
commonly is separated from the full amplitude.
  We derive a general expression for the structure-dependent terms in
an expansion in terms of the momenta $k_1$ and $k_2$ of the initial
and final virtual photon, respectively, up to and including terms of order
${\cal O}(k^4)$.
   At order ${\cal O}(k^2)$ two terms appear which are related to the
usual electric and magnetic polarizabilities of real Compton scattering.
   At order ${\cal O}(k^4)$ we find nine new structures of which
five can only be determined using virtual photons.
\end{abstract}

\section{Introduction }
\label{intro}

   It has been known for a long time that the two leading orders of an
expansion of the real Compton scattering amplitude off the proton in
terms of the frequency of the photon are determined by a low-energy
theorem (LET) \cite{Low1,GellMann}:
the coefficients are entirely given in terms of the charge, mass and
anomalous magnetic moment of the proton.
   This theorem is based on Lorentz invariance, gauge invariance and
crossing symmetry.
   Only at second order in the frequency can one test new model-dependent
structures, namely, the electric and magnetic polarizabilities
(for an overview see, e.g., \cite{Lvov}).

   Recently, the process of virtual Compton scattering off a nucleon has
attracted considerable interest from both the experimental
\cite{Audit1,Brand,Breton,Audit2} and the theoretical side
\cite{Schaefer,Guichon,Vanderhaeghen,Scherer1,Liu,Metz,Hemmert1,Hemmert2}.
   In contrast to the scattering of real photons, the use of virtual photons
allows to independently vary the energy and the momentum transferred to or
from the target.
   Thus the possibilities of investigating the structure of the nucleon
are increased tremendously as compared with real Compton scattering.
   The extension of the LET to include virtual photons as well was given in
\cite{Guichon,Scherer1}.
   Furthermore, the structure dependence beyond the LET was studied in detail
in \cite{Guichon}.
   For example, the reaction $e+p\to e+p+\gamma$ involving a low-energy
photon in the final state is described in terms of ten generalized
polarizabilities \cite{Guichon} as opposed to two electromagnetic
polarizabilities in real Compton scattering $\gamma +p\to \gamma +p$.

   Here, we want to consider virtual Compton scattering off a scalar or
pseudoscalar particle.
   In this context one might either think of ``stable'' elementary particles,
such as the pion or kaon, but also of spin-zero nuclei.
   By limiting ourselves to a spin-zero process we hope to be able to
understand the basic features without getting lost in the algebra necessary
to handle spin as in the perhaps more easily measurable Compton scattering off
a nucleon.
   In principle, one can distinguish between three different cases:
the virtual Compton scattering off (a) a charged particle, e.g., a
$\pi^+$ or $K^+$; (b) a neutral particle which is
not its own antiparticle, e.g., a $K^0$;
and (c) a neutral particle which is its own antiparticle, e.g., a $\pi^0$.
   In this work we will focus on the first situation since this is the one
where the Ward-Takahashi identities \cite{Ward,Takahashi} provide
non-trivial information.
   The second and third cases are covered by our discussion if one turns
off the constraints of the Ward-Takahashi relations.
   Due to charge conjugation invariance a single photon cannot couple to
a particle which is its own antiparticle.
   This distinguishes (b) from (c) in that there are no pole terms in
the Compton scattering off a neutral pion.

   The aim of this work is to use an investigation of, say, virtual Compton
scattering off a $\pi^+$ to address two major questions.
   First, what is the difference between the usual soft-photon approximation
(SPA) approach \cite{Low2,AdDot} to deriving a low-energy expression and the
alternative approach \cite{GellMann,Kazes} using the Ward-Takahashi (WT)
relations.
   Second, to what extent is the low-energy amplitude so derived model
independent and what can be said about the general structure of the
leading model-dependent terms.

\section{Framework}
\label{frame}

   In order to be specific we consider the virtual Compton scattering process
$\gamma(k_1) + \pi^+(p_i)\rightarrow \gamma(k_2) + \pi^+(p_f)$
for four-momenta satisfying $k_1 + p_i = k_2 + p_f$.
   We will always assume that the photons can be virtual, i.e., $k_1^2$
and $k_2^2$ are non-zero.
   Amplitudes (or pion legs) will be described as ``on shell'' if
$p_i^2 = m^2$ and $p_f^2 = m^2$, where $m$ is the mass of the pion,
and ``off shell'' otherwise.
   Thus the concepts of on and off shell will apply to the pions.

   The most general irreducible, renormalized, photon-pion-pion vertex can
then be written as \cite{Nishijima,Rudy}
\begin{equation}\label{genver}
\Gamma^{\mu}(p\prime,p)=F(k^2,{p\prime}^2,p^2)(p\prime + p)^{\mu} +
G(k^2,{p\prime}^2,p^2)(p\prime - p)^{\mu},
\end{equation}
where $k=p\prime - p$ and $F$ and $G$ are form functions, which are taken to be
functions of the squares of the four-momenta of the photon and pions.
   In the following, we will distinguish between form functions and form
factors, where the former may depend on the representation chosen and only
the latter necessarily correspond to observable quantities.

   We will assume that the vertex satisfies the usual Ward-Takahashi relation
\cite{Ward,Takahashi}
\begin{equation} \label{vertex}
k_{\mu} \Gamma^{\mu}(p\prime,p) = \Delta^{-1}(p\prime)-\Delta^{-1}(p),
\end{equation}
where $\Delta (p)$ is the full renormalized pion propagator which satisfies
\cite{appendixRudy}
\begin{equation}
\Delta^{-1}(p) = (p^2-m^2)F(0,p^2,m^2) =(p^2-m^2)F(0,m^2,p^2).
\end{equation}
   This allows us to write for $G$ \cite{appendixRudy}
\begin{equation}\label{gffact}
G(k^2,{p\prime}^2,p^2) = \frac{({p\prime}^2 - p^2)}{k^2}[F(0,{p\prime}^2,p^2)
- F(k^2,{p\prime}^2,p^2)].
\end{equation}

   For some applications it will be convenient to adopt a simplified form of
this general vertex function in which the off-shell dependence of
$F(k^2,{p\prime}^2,p^2)$ is neglected.
   In this representation
\begin{eqnarray} \label{nooff}
F(k^2,{p\prime}^2,p^2)& \rightarrow & F(k^2,m^2,m^2) \equiv F(k^2),\nonumber\\
G(k^2,{p\prime}^2,p^2)& \rightarrow &
\frac{({p\prime}^2 - p^2)}{k^2} [1 - F(k^2)].
\end{eqnarray}
   The rationale for this assumption arises from the fact that one can make
transformations of the fields in an effective Lagrangian, essentially
redefinitions of the fields, which do not change the physical observable
quantities but which allow to a certain extent to transform away the off-shell
dependence at the vertices
\cite{Lvov,Chisholm,Kamefuchi,Arzt,Scherer2,Scherer3}.
   Thus the full amplitude, which is the measurable quantity, should
not be affected by this particular choice.
   For now, however, we continue to use the general form Eqs.\ (\ref{genver}),
(\ref{gffact}) and will adopt the simplified form only when explicitly
specified.

   Using these definitions we can write the matrix element for the two pole
diagrams, direct and crossed virtual photon scattering off pions, as
\begin{equation}
M^{\mu\nu}_{pole} = \Gamma^{\nu}(p_f,p_f+k_2) \Delta(p_f+k_2)
\Gamma^{\mu}(p_i+k_1,p_i)
+ \Gamma^{\mu}(p_f,p_f-k_1) \Delta(p_f-k_1) \Gamma^{\nu}(p_i-k_2,p_i),
\end{equation}
where we have dropped an overall factor of $-ie^2$, with $e > 0$ the electric
charge, $e^2/4\pi\approx 1/137$, arising from the usual Feynman rules
\cite{Bjorken}.
   The contribution to the physical matrix element arises from, of course, the
on-shell limit of this expression.

   This pole contribution, $M^{\mu\nu}_{pole}$, is the analog of the
``radiation from external lines'' contributions used as a starting point
in soft-photon expansions.
   It contains all the contributions which are non-analytic as either
$k_1 \rightarrow 0$ or $k_2 \rightarrow 0$.
   There will, of course, be many other contributions to the full matrix
element including those from seagull-type diagrams or from intermediate
resonances.
   All of these other contributions are non-singular in the limit $k_1
\rightarrow 0$ or $k_2 \rightarrow 0$, however.

\section{Comparison of SPA and WT}
\label{comp}

   In this section we want to compare in detail the usual approach of the
soft-photon approximation (SPA) and the analogous Ward-Takahashi (WT) approach
to deriving the low-energy behavior for this process.
   For both approaches it will be useful to divide the full amplitude into
two pieces so that
\begin{equation}
M^{\mu \nu} = M^{\mu \nu}_A +  M^{\mu \nu}_B.
\end{equation}
   Here $M^{\mu \nu}_A$ will contain all of the terms in the amplitude which
are singular as either $k_1 \rightarrow 0$ or $k_2 \rightarrow 0$, together,
perhaps, with some additional non-singular terms.
   $M^{\mu \nu}_B$ will contain everything else.
   We stress that this separation is not unique in the sense that
non-singular terms may be shifted from $M^{\mu\nu}_A$ to $M^{\mu\nu}_B$ and
vice versa.

   We shall see that the conditions imposed on the full amplitude in the two
approaches are exactly the same on shell, so that the physical results from the
two approaches must be exactly the same.
   However, the natural way to divide the full amplitude into $M^{\mu \nu}_A$
and $M^{\mu \nu}_B$ will differ.
   Also one can, in principle, extend the WT approach more easily to
considerations of off-shell amplitudes.

\subsection{Soft-Photon Approximation}
\label{ssSPA}

   In the usual soft-photon approach to radiative processes
\cite{Low2,AdDot} one starts with the amplitude for radiation from external
legs, which contains all of the $1/k$ singularities, and expands this
amplitude in the explicit $k$ dependence about $k=0$, keeping terms of
${\cal O}(1/k)$ and ${\cal O}(k/k)$.
   Gauge invariance is then used to fix the ${\cal O}(k^0)$ terms which are
independent of $k$.
   Higher-order terms are then not determined.
   There are also usually a variety of ways of defining the on-shell
information in the pole terms and of making the expansions, and one can
always show that different ways lead to results which differ only by terms of
${\cal O}(k)$ or higher.

   For a process involving virtual photons, and more than one of them,
this approach needs a slight generalization.
   We first start with the amplitude for the pole graphs $M^{\mu\nu}_{pole}$,
taken on shell, as these are the only graphs which contain singularities as
$k_1$ or $k_2$ tend to zero.
   The ``expansion in explicit $k$ dependence'' becomes an expansion of the
vertex functions about the on-shell points.
   In processes with real photons one is actually doing the same thing.
   However, there $k$ appears only in the combinations like $p_f \pm k$ and
an expansion in explicit $k$ dependence is the same as an expansion about the
on-shell point.
   Here, since the photons are virtual, $k$ appears also as $k^2$ in
the form functions and form factors.
   We explicitly do not expand such form factors about $k=0$.
   If we did there would be terms of arbitrarily high order in $k$ which
would, however, have $k$ also in the denominator, and thus we would not have
isolated all the singular terms.

   When we expand the vertex function $\Gamma^\mu$ we see that the form
function $F$ becomes simply $F(k^2)$ plus terms proportional to
$\Delta^{-1}_F(p) = (p^2-m^2)$, where $\Delta_F(p)$ is the free Feynman
propagator.
   The $\Delta^{-1}_F(p)$ cancels the similar factor in the denominator and
leads to a term which is non-singular.
   Similarly the $G$ form function is directly proportional to
$\Delta_F^{-1}(p)$ and so also generates a non-singular term.
   The result, which we take as $M^{\mu\nu}_{A-SPA}$, is
\begin{equation}
M^{\mu\nu}_{A-SPA} = F(k_1^2) F(k_2^2)\left [
\frac{(2p_i+k_1)^\mu (2p_f+k_2)^\nu}{s-m^2}
+ \frac{(2p_f-k_1)^\mu (2p_i-k_2)^\nu}{u-m^2} \right ],
\end{equation}
where $s=(p_i+k_1)^2=(p_f+k_2)^2$ and $u=(p_i-k_2)^2=(p_f-k_1)^2$.

   The next step is to apply the gauge condition to the full amplitude
$M^{\mu\nu}_{SPA} =M^{\mu\nu}_{A-SPA}+M^{\mu\nu}_{B-SPA}$.
   Thus we have for the full on-shell amplitude
\begin{equation} \label{gauge}
{k_1}_\mu M^{\mu\nu} = 0,\quad M^{\mu\nu}{k_2}_\nu = 0 .
\end{equation}
   This gives in this case ${k_1}_\mu M^{\mu\nu}_{A-SPA} =
2F(k_1^2)F(k_2^2)k_1^\nu$, $M^{\mu\nu}_{A-SPA}{k_2}_\nu =
2F(k_1^2)F(k_2^2)k_2^\mu$ and thus $M^{\mu\nu}_{B-SPA} =
-2F(k_1^2)F(k_2^2)g^{\mu\nu}$.
   The resulting full soft-photon amplitude is then
\begin{equation}
\label{fspa}
M^{\mu\nu}_{SPA} = F(k_1^2) F(k_2^2)\left [
\frac{(2p_i+k_1)^\mu (2p_f+k_2)^\nu}{s-m^2}
+ \frac{(2p_f-k_1)^\mu (2p_i-k_2)^\nu}{u-m^2} -2g^{\mu\nu} \right ] +
\tilde{M}^{\mu\nu}_{B-SPA}.
\end{equation}
   The last term, $\tilde{M}^{\mu\nu}_{B-SPA}$, is gauge invariant by itself,
i.e., satisfies Eq.\ (\ref{gauge}).
   One can always add such terms to an amplitude which is fixed only by gauge
invariance and must do so to obtain a completely general result.

   Observe that the $M^{\mu\nu}_{A-SPA}+M^{\mu\nu}_{B-SPA}$ part of this
amplitude is gauge invariant and contains all of the terms which are singular
as either $k_1 \rightarrow 0$ or $ k_2 \rightarrow 0$.
   It satisfies crossing symmetry ($k_1 \leftrightarrow -k_2$,
$\mu \leftrightarrow \nu$) as it must.
   It also is completely determined by a knowledge of the on-shell
information, i.e., by the on-shell form factor and charge.
   It is unique through  ${\cal O}(k^0)$ in the sense that any other
expression satisfying these conditions can differ only by terms of
${\cal O}(k)$ or higher.
   Likewise information from unknown contributions appears first at
${\cal O}(k)$.
   In fact, when discussing the structure dependence, we shall see that
even the ${\cal O}(k)$ terms are determined by Eq.\ (\ref{fspa}) which
can be interpreted as a result of applying two gauge invariance conditions
as opposed to one in the usual soft-photon approximation, or alternatively
as applying gauge invariance once and then imposing crossing symmetry.

   In this SPA approach $M^{\mu\nu}_{A-SPA}$ represents in some sense a
minimal singular term.
   It contains only the singular pieces of $M^{\mu\nu}_{pole}$, which is then
made gauge invariant by adding $M^{\mu\nu}_{B-SPA}$.
   All other terms are included in $\tilde{M}^{\mu\nu}_{B-SPA}$.

\subsection{Ward-Takahashi Approach}
\label{ssWT}

   We now want to consider an alternative approach \cite{GellMann,Kazes}
which is based on the use of the Ward-Takahashi relations.
   In this approach we start also with the most general $M^{\mu\nu}_{pole}$
and again develop an amplitude divided into two parts $M^{\mu\nu}_{A-WT}$ and
$M^{\mu\nu}_{B-WT}$ with all singular contributions included in
$M^{\mu\nu}_{A-WT}$.
   We then impose on the vertex and the full amplitude the conditions
\cite{GellMann,Kazes}:
\begin{equation}
k_{\mu} \Gamma^{\mu}(p\prime,p) = \Delta^{-1}(p\prime)-\Delta^{-1}(p),
\label{WT1}
\end{equation}
\begin{equation}
{k_1}_\mu M^{\mu\nu} = \Delta^{-1}(p_f) \Delta(p_i-k_2) \Gamma^\nu(p_i-k_2,p_i)
-\Gamma^\nu(p_f,p_f+k_2)\Delta(p_f+k_2)\Delta^{-1}(p_i), \label{WT2}
\end{equation}
\begin{equation}
M^{\mu\nu}{k_2}_\nu =- \Delta^{-1}(p_f) \Delta(p_i+k_1) \Gamma^\mu(p_i+k_1,p_i)
+\Gamma^\mu(p_f,p_f-k_1)\Delta(p_f-k_1)\Delta^{-1}(p_i). \label{WT3}
\end{equation}
   The third of these equations is obtained from the second by crossing,
($k_1 \leftrightarrow -k_2$, $\mu \leftrightarrow \nu$), and the first is
just the condition Eq.\ (\ref{vertex}) imposed also on the vertex function
used in the SPA approach.
   In principle these equations hold also off shell.
   However, the last two apply to the full amplitude, which is really an
observable only on shell.
   In the on-shell limit $\Delta^{-1} \rightarrow 0$ and the two conditions
reduce to the same conditions, those of Eq.\ (\ref{gauge}), used in the SPA.
   One in addition must ensure that crossing symmetry is obeyed.

   We see then that the conditions imposed both on the amplitude and on the
vertex are exactly the same in the WT approach as in the SPA approach, as
long as one considers the physically relevant on-shell limit for the full
amplitude.
   Thus we conclude that the unique information about the full amplitude
resulting from these two approaches must be the same.
   We will see, however, that the natural division into amplitudes
$M^{\mu\nu}_{A-WT}$ and $M^{\mu\nu}_{B-WT}$ is somewhat different than it was
in the SPA case.

   To start take $M^{\mu\nu}_{A-WT} = M^{\mu\nu}_{pole}$.
   It thus contains the complete pole terms, including in particular all of
the singularities.
   As compared to the choice for $M^{\mu\nu}_{A-SPA}$, it includes many
higher-order terms, and contributions from the off-shell dependence of the
vertex function and the full propagator.
   However, we have to keep in mind that these ``off-shell effects''
depend on the representation one chooses and thus are not uniquely
determined.

   Now, taking $M^{\mu\nu}_{WT} = M^{\mu\nu}_{A-WT} + M^{\mu\nu}_{B-WT}$,
we contract with $k_{1\mu}$ and use the condition of Eq.\ (\ref{WT2}), for
the moment considering the full off-shell amplitude and using the condition
off shell.
   This gives
\begin{eqnarray}
{k_1}_\mu M^{\mu\nu}_{WT} &=&
\Delta^{-1}(p_f) \Delta(p_i-k_2) \Gamma^\nu(p_i-k_2,p_i)
-\Gamma^\nu(p_f,p_f+k_2)\Delta(p_f+k_2)\Delta^{-1}(p_i) \nonumber \\
&=& \Gamma^{\nu}(p_f,p_f+k_2) \Delta(p_f+k_2) {k_1}_\mu
\Gamma^{\mu}(p_i+k_1,p_i) \nonumber \\
&& + {k_1}_\mu \Gamma^{\mu}(p_f,p_f-k_1) \Delta(p_f-k_1)
\Gamma^{\nu}(p_i-k_2,p_i) + {k_1}_\mu M^{\mu\nu}_{B-WT}.
\end{eqnarray}
   Eq.\ (\ref{WT1}) is then used to express the $k \cdot \Gamma$'s in terms of
$\Delta^{-1}$'s and we get as a condition on $M^{\mu\nu}_{B-WT}$
\begin{eqnarray}
{k_1}_\mu M^{\mu\nu}_{B-WT} &=& \Gamma^\nu(p_i-k_2,p_i) -
\Gamma^\nu(p_f,p_f+k_2), \label{kdotm1} \\
M^{\mu\nu}_{B-WT} {k_2}_\nu &=& -\Gamma^\mu(p_i+k_1,p_i) +
\Gamma^\mu(p_f,p_f-k_1), \label{kdotm2}
\end{eqnarray}
where the second of these can be obtained by direct calculation or by
crossing from the first.
   Note that while we formally derived these relations by considering
off-shell amplitudes, we would have obtained exactly the same
conditions on $M^{\mu\nu}_{B-WT}$ had we worked entirely with on-shell
quantities.
   The conditions on $M^{\mu\nu}_{B-WT}$ are different from those
obtained for $M^{\mu\nu}_{B-SPA}$ because the initial choice for
$M^{\mu\nu}_{A-WT}$ is different than that for $M^{\mu\nu}_{A-SPA}$.

   The next step is to determine $M^{\mu\nu}_{B-WT}$, which it turns out is
not as easy as it was for the SPA case, as one cannot easily guess at a
solution, but must solve explicitly.
   First define $P^\mu = p_f^\mu +p_i^\mu$ and consider
$P^\mu$, $k_1^\mu$, and $k_2^\mu$ as the three independent four-vectors.
Let
\begin{eqnarray}
F_{1f}&=&F(k_1^2,p_f^2,(p_f-k_1)^2), \nonumber \\
F_{2f}&=&F(k_2^2,p_f^2,(p_f+k_2)^2), \nonumber \\
F_{1i}&=&F(k_1^2,(p_i+k_1)^2,p_i^2), \nonumber \\
F_{2i}&=&F(k_2^2,(p_i-k_2)^2,p_i^2),  \label{fdef}
\end{eqnarray}
with a similar set of definitions for the $G(k^2,{p\prime}^2,p^2)$ form
functions.

   With these definitions Eqs.\ (\ref{kdotm1}), (\ref{kdotm2}) become
\begin{eqnarray}
{k_1}_\mu M^{\mu\nu}_{B-WT} &=& (F_{2i}-F_{2f})P^\nu -(F_{2i}+F_{2f})k_1^\nu
+(G_{2f}-G_{2i})k_2^\nu,  \label{k1m} \\
M^{\mu\nu}_{B-WT} {k_2}_\nu &=& -(F_{1i}-F_{1f})P^\mu -(F_{1i}+F_{1f})k_2^\mu
+(G_{1f}-G_{1i})k_1^\mu . \label{k2m}
\end{eqnarray}

   To actually determine $M^{\mu\nu}_{B-WT}$ we start with the most general
second-rank tensor which can be written with the metric tensor and
the three independent four-vectors we have,\footnote{The
completely antisymmetric pseudotensor $\epsilon^{\alpha\beta\gamma\delta}$
is excluded due to parity.} namely,
\begin{equation}
M^{\mu\nu}_{B-WT} \sim
 g^{\mu\nu}, P^\mu P^\nu, P^\mu k_1^\nu, k_1^\mu P^\nu, P^\mu k_2^\nu,
k_2^\mu P^\nu, k_1^\mu k_1^\nu, k_2^\mu k_2^\nu, k_1^\mu k_2^\nu,
k_2^\mu k_1^\nu,
\end{equation}
and substitute it into the left-hand side of Eqs.\ (\ref{k1m}), (\ref{k2m})
above.
   Matching the coefficients of the independent four-vectors superficially
leads to six equations, out of which only five are linearly independent.
    These can be solved for the coefficients in the general form for
$M^{\mu\nu}_{B-WT}$.
   We find then, having dropped for now those terms which are separately gauge
invariant and which do not depend on the form functions $F$ or $G$,
\begin{eqnarray}
M^{\mu\nu}_{B-WT1}= &-&\frac{1}{2}g^{\mu\nu}(F_{1i}+F_{1f}+F_{2i}+F_{2f})
-(F_{1i}-F_{1f})\frac{P^\mu k_2^\nu}{k_2^2}
+(F_{2i}-F_{2f})\frac{k_1^\mu P^\nu}{k_1^2} \nonumber \\
&+&\frac{1}{k_1^2 k_1 \cdot k_2}[k_1^2(G_{1f}-G_{1i})-(F_{2i}-F_{2f})
k_2 \cdot P]\left(k_1^\mu k_1^\nu - \frac{k_1^2 g^{\mu\nu}}{2}\right)
\nonumber \\
&+&\frac{1}{k_2^2 k_1 \cdot k_2}[k_2^2(G_{2f}-G_{2i})+(F_{1i}-F_{1f})
k_1 \cdot P]\left(k_2^\mu k_2^\nu - \frac{k_2^2 g^{\mu\nu}}{2}\right).
\label{mbwt1}\end{eqnarray}
   We have written this in an explicitly crossing-symmetric way valid also off
shell, although on shell we have $k_1 \cdot P = k_2 \cdot P$.
   It has also been labeled by WT1 to distinguish it from a second form of
the WT result to be considered shortly.
   In the same notation
\begin{eqnarray}
M^{\mu\nu}_{A-WT1}&=& \frac{[F_{1i}(2p_i+k_1)^\mu + G_{1i} k_1^\mu]
[F_{2f}(2p_f+k_2)^\nu - G_{2f} k_2^\nu]}{(s-m^2)F(0,m^2,(p_f+k_2)^2)}
\nonumber \\
&&+\frac{[F_{1f}(2p_f-k_1)^\mu + G_{1f} k_1^\mu]
[F_{2i}(2p_i-k_2)^\nu - G_{2i} k_2^\nu]}{(u-m^2)F(0,m^2,(p_f-k_1)^2)},
\label{mawt1}\end{eqnarray}
and the full amplitude is
\begin{equation}
M^{\mu\nu}_{WT1} = M^{\mu\nu}_{A-WT1} + M^{\mu\nu}_{B-WT1} +
\tilde{M}^{\mu\nu}_{B-WT1},
\end{equation}
where again $\tilde{M}^{\mu\nu}_{B-WT1}$ is an undetermined piece which is
separately gauge invariant, and where so far it has not been necessary to
specialize to the on-shell case.

   The full amplitude $M^{\mu\nu}_{WT1}$ is of course the same on shell as
the full SPA amplitude $M^{\mu\nu}_{SPA}$ as we are simply expressing the
same result in different ways.
   However, each of the three individual pieces is different.
   As compared to SPA the $M^{\mu\nu}_{A-WT1}$ obtained is in some sense a
maximal singular term.
   It consists of the complete pole term and thus contains all the
singular terms but also a large number of higher-order non-singular terms,
which in the SPA were lumped into $\tilde{M}^{\mu\nu}_{B-SPA}$.
   It contains no more real information, however, as there are still unknown
terms of ${\cal O}(k)$ or higher which are not included.
   The combination $M^{\mu\nu}_{A-WT1} + M^{\mu\nu}_{B-WT1}$ differs from the
similar form for the SPA approach by terms which are separately gauge
invariant and of ${\cal O}(k)$ or higher.
   It contains information from the on-shell form factors, but also in
principle, unlike the analogous SPA term, contributions from the off-shell
dependence of the form functions.

   We now want to consider a third possibility, which can be obtained from
$M^{\mu\nu}_{WT1}$ by making the specific assumption discussed earlier that
the form functions have no off-shell dependence.
   Thus we substitute the specific form of Eq.\ (\ref{nooff}) into
Eqs.\ (\ref{mbwt1}), (\ref{mawt1}) above and take the on-shell limit to
obtain a second WT amplitude, namely,
\begin{eqnarray}
M^{\mu\nu}_{A-WT2} &=& F(k_1^2) F(k_2^2)
\left [ \frac{(2p_i+k_1)^\mu (2p_f+k_2)^\nu}{s-m^2}
+ \frac{(2p_f-k_1)^\mu (2p_i-k_2)^\nu}{u-m^2} \right ] \nonumber \\
&&+ 2F(k_1^2) \frac{[1-F(k_2^2)]}{k_2^2} k_2^\mu k_2^\nu
+ 2F(k_2^2) \frac{[1-F(k_1^2)]}{k_1^2} k_1^\mu k_1^\nu \nonumber \\
&&+ 2k_1 \cdot k_2 \frac{[1-F(k_1^2)]}{k_1^2} \frac{[1-F(k_2^2)]}{k_2^2}
k_1^\mu k_2^\nu,
\end{eqnarray}
and
\begin{equation}
M^{\mu\nu}_{B-WT2} =2g^{\mu\nu}[1-F(k_1^2)-F(k_2^2)]
-2 \frac{[1-F(k_1^2)]}{k_1^2} k_1^\mu k_1^\nu
-2 \frac{[1-F(k_2^2)]}{k_2^2} k_2^\mu k_2^\nu,
\end{equation}
with, as before,
\begin{equation}
M^{\mu\nu}_{WT2} = M^{\mu\nu}_{A-WT2} + M^{\mu\nu}_{B-WT2}
+ \tilde{M}^{\mu\nu}_{B-WT2}.
\end{equation}
   This particular choice has the advantage that $M^{\mu\nu}_{A-WT2} +
M^{\mu\nu}_{B-WT2}$ depends only on on-shell information, and yet keeps all of
the singular terms and some of the higher-order terms of the pole contribution.
   One can see more clearly the comparison with the SPA amplitude by adding and
subtracting from $M^{\mu\nu}_{A-WT2}$ the  quantity
$-2F(k_1^2)F(k_2^2)g^{\mu\nu}$ which puts it in the same form as the SPA
amplitude.
   Thus one gets
\begin{eqnarray}
M^{\mu\nu}_{WT2} &=& M^{\mu\nu}_{A-WT2} + M^{\mu\nu}_{B-WT2}
+ \tilde{M}^{\mu\nu}_{B-WT2} \nonumber \\
&=& M^{\mu\nu}_{A-SPA} + M^{\mu\nu}_{B-SPA}
+ \tilde{M}^{\mu\nu}_{B-WT2} \nonumber \\
&&- 2\frac{[1-F(k_1^2)-F(k_2^2)+F(k_1^2)F(k_2^2)]}{k_1^2k_2^2}
[k_1^2 k_2^\mu k_2^\nu +k_2^2 k_1^\mu k_1^\nu - k_1 \cdot k_2 k_1^\mu k_2^\nu
-k_1^2 k_2^2 g^{\mu\nu}].\nonumber\\
\end{eqnarray}
   Note that the factor $[1-F(k_1^2)-F(k_2^2)+F(k_1^2)F(k_2^2)]$ is actually
proportional to $k_1^2 k_2^2$ when each form factor
is expanded in $k^2$ and so this term is not really singular.
   From this equation it is clear that $M^{\mu\nu}_{WT2}$ differs from
$M^{\mu\nu}_{SPA}$ simply in that a piece which is gauge invariant by itself,
and in this case of ${\cal O}(k^4)$, has been pulled out of
$\tilde{M}^{\mu\nu}_{B-SPA}$ to give $\tilde{M}^{\mu\nu}_{B-WT2}$ and moved
over to $M^{\mu\nu}_{A-WT2} + M^{\mu\nu}_{B-WT2}$.

\subsection{Summary of Comparison}
\label{sscompsum}

   We have developed three different low-energy expressions, one using the SPA
approach and two originating in variations of the WT approach.
   Clearly there could be many other variations.
   Thus we want to summarize here what seem to be the general features of
these various approaches and the ways in which they differ.

   The first observation is that the conditions imposed on the full on-shell
amplitude for the virtual Compton scattering process in the SPA approach and in
the WT approach are essentially the same and are those given by
Eqs.\ (\ref{gauge}) together with crossing symmetry and Eq.\ (\ref{vertex})
for the vertex.
   The WT conditions can in principle be extended off shell, but when
applied to the physically relevant on-shell matrix element, they lead to the
same conditions as for SPA.
   This means that the physical content of the amplitudes derived in the two
approaches must be exactly the same.

   We did see, however, that the natural way to divide the amplitudes up into
pieces, $M^{\mu\nu}=M^{\mu\nu}_A +M^{\mu\nu}_B + \tilde{M}^{\mu\nu}_B$
differs in the various approaches, though in all cases all terms singular as
either $k_1 \rightarrow 0$ or $k_2 \rightarrow 0$ are kept in $M^{\mu\nu}_A$.
   In the SPA approach the combination $M^{\mu\nu}_A +M^{\mu\nu}_B$ contains
the minimal set of terms, i.e., just the pole terms and the minimal factors
needed to make them gauge invariant.
   Everything else is pushed into $\tilde{M}^{\mu\nu}_B$.
   This approach has the advantage of giving an essentially unique result, but
the disadvantage of pushing some higher-order pieces, which are in fact known,
into the unknown $\tilde{M}^{\mu\nu}_B$.

   In the WT1 approach $M^{\mu\nu}_A +M^{\mu\nu}_B$ contains the full
contribution of the most general pole terms to all orders, together with
enough factors extracted from $\tilde{M}^{\mu\nu}_B$ to make the amplitude
gauge invariant.
   This approach has the advantage of incorporating explicitly all information
present in the pole terms.
   However, the pieces needed to make the pole terms gauge invariant come in
principle from other unknown diagrams, so this mixes physics from these unknown
diagrams with the known parts, and it is not clear whether this is better or
worse.
   This approach also has the disadvantage of using explicitly off-shell form
functions, which we know cannot in the end be uniquely determined from the
total amplitude.

   Finally, in the third approach, labeled by WT2, we drop any off-shell
dependence of the form functions in $M^{\mu\nu}_A +M^{\mu\nu}_B$.
   This has the advantage now of requiring only on-shell information to
evaluate the leading terms, but otherwise has the same advantages and
disadvantages of WT1.

   Formally these different approaches differ from one another primarily in
the different higher-order terms which they remove from
$\tilde{M}^{\mu\nu}_B$ and put into the pole terms.
   Such terms are at least of ${\cal O}(k)$ and are gauge invariant by
themselves, and so cannot be uniquely fixed by conditions depending on gauge
invariance alone.

   The really important observations here, however, are that there are
multiple low-energy expressions, depending on how the various terms are
apportioned.
   Furthermore, different choices for the leading terms imply different
choices for $\tilde{M}^{\mu\nu}_B$.
   Since it is this piece which contains structure-dependent terms such as
polarizabilities this means that the definition of these quantities must
specify also what form is used for the leading terms.
   One cannot simply determine these structure terms in isolation.

\section{Structure Terms}
\label{struct}

   In this section we want to address the last question posed in the
introduction, namely what can be said in general about the form of the
leading model-dependent or structure terms, i.e., can we put any constraints
on the form of $\tilde{M}^{\mu\nu}_B$.

   The first observation is that the only constraints we have, apart from
Lorentz invariance and the discrete symmetries, are those coming from gauge
invariance and crossing symmetry and from the fact that all singularities
have been extracted so that $\tilde{M}^{\mu\nu}_B$ must be finite as either
$k_1 \rightarrow 0$ or $k_2 \rightarrow 0$.
   Thus we must have
\begin{equation} \label{gaugeB}
k_{1\mu} \tilde{M}^{\mu\nu}_B = 0,\quad \tilde{M}^{\mu\nu}_B k_{2\nu} = 0,
\end{equation}
and in addition $\tilde{M}^{\mu\nu}_B$ must be crossing symmetric, since the
leading terms have been explicitly made crossing symmetric.
   Note that these constraints are the same regardless of which of the above
choices for the leading terms is made, as long as we consider on-shell
amplitudes.
   Thus the general form we derive will be the same in all cases.
   However, as emphasized above, the specific value of the coefficients will
depend on the choice of the leading term.
   For the WT1 and WT2 cases these constraints hold for off-shell
amplitudes also, since we have already satisfied the full off-shell WT
identities of Eqs.\ (\ref{WT2}), (\ref{WT3}) with the
$M^{\mu\nu}_A+M^{\mu\nu}_B$
parts.
   However, we will for this section work just on shell and only at the end
briefly comment on modifications for the off-shell case.

   We start with the general form of $\tilde{M}^{\mu\nu}_B$ written in the form
\begin{eqnarray}
\tilde{M}^{\mu\nu}_B &=& A g^{\mu\nu} + B P^\mu P^\nu
 +  C(P^\mu k_1^\nu-k_2^\mu P^\nu) + \tilde{C}(P^\mu k_1^\nu+k_2^\mu P^\nu)
\nonumber \\
&&+ D(P^\mu k_2^\nu-k_1^\mu P^\nu) + \tilde{D}(P^\mu k_2^\nu+k_1^\mu P^\nu)
 +  E(k_1^\mu k_1^\nu + k_2^\mu k_2^\nu)
+ \tilde{E}(k_1^\mu k_1^\nu - k_2^\mu k_2^\nu) \nonumber \\
&&+ F k_1^\mu k_2^\nu + G k_2^\mu k_1^\nu,
\end{eqnarray}
where $\tilde{C},\tilde{D}$, and $\tilde{E}$ are odd under crossing whereas the
other coefficients are even.
   The coefficients are functions of the independent scalars which can be
formed from the three independent four-vectors $k_1, k_2, P$,
which we take to be $k_1 \cdot k_2$ and $(k_1^2+k_2^2)$ which are even
under crossing and $(k_1^2-k_2^2)$ and $(k_1+k_2) \cdot P$ which are odd.
   The remaining two scalar variables are $(k_1-k_2) \cdot P = p_f^2-p_i^2$
and $P^2 = 2(p_f^2+p_i^2)+2k_1 \cdot k_2 -k_1^2 -k_2^2$. On shell, with $p_i^2
=p_f^2=m^2$, the first of these is zero and the second is not independent of
the others kept and so both can be dropped.

   Parity is already taken care of, whereas the symmetry with respect
to a charge conjugation transformation implies that the amplitude is unchanged
under the exchange $p_i \leftrightarrow -p_f$ which, in terms of the variables
we use, gives
\begin{equation}\label{csym}
\tilde{M}^{\mu\nu}_B(P,k_1,k_2)=\tilde{M}_B^{\mu\nu}(-P,k_1,k_2).
\end{equation}
   In other words, the coefficients $A$, $B$, $E$, $\tilde{E}$, $F$, $G$
are even under charge conjugation and the other coefficients are odd.
   On shell, only $(k_1+k_2)\cdot P$ receives a minus sign under charge
conjugation whereas the other scalar variables remain unchanged.
   Off shell, also $(k_1-k_2)\cdot P$ is odd under $C$.

   Time reversal invariance implies in this case that the full amplitude is
invariant under the simultaneous interchange $p_i, k_1, \epsilon_1
\leftrightarrow p_f, k_2, \epsilon_2$ or equivalently that
$\tilde{M}^{\mu\nu}_B$ is invariant under $p_i, k_1, \mu
\leftrightarrow p_f, k_2, \nu$. This requires that the coefficients $C, D,
\tilde{E}$ be odd and the others even under this transformation. The scalar
variables $(k_1^2-k_2^2)$ and $(k_1-k_2) \cdot P$ are odd and the others
are even.

   We then impose the gauge conditions of Eq.\ (\ref{gaugeB}).
   Since the four-vectors $P^\mu,k_1^\mu,k_2^\mu$ are independent their
coefficients must vanish in these conditions and we are led to the following
six equations:
\begin{eqnarray}\label{geqns}
0 &=& A+(C+\tilde{C})k_1 \cdot P+(E+\tilde{E})k_1^2+Gk_1 \cdot k_2,\nonumber\\
0 &=& Bk_1 \cdot P-(C-\tilde{C})k_1 \cdot k_2-(D-\tilde{D})k_1^2,\nonumber\\
0 &=& (D+\tilde{D}) k_1 \cdot P+(E-\tilde{E})k_1 \cdot k_2+Fk_1^2,\nonumber\\
0 &=& A-(C-\tilde{C})k_2 \cdot P+(E-\tilde{E})k_2^2+Gk_1 \cdot k_2,\nonumber\\
0 &=& Bk_2 \cdot P+(C+\tilde{C})k_1 \cdot k_2+(D+\tilde{D})k_2^2,\nonumber\\
0 &=& -(D-\tilde{D}) k_2 \cdot P+(E+\tilde{E})k_1 \cdot k_2+Fk_2^2.
\end{eqnarray}
   As noted before, only five of these equations are linearly independent.

   Next we expand each of the coefficients in the form
$A=A_0+A_1+A_2+A_3+A_4+\cdots$, where the subscript refers to the power of $k$
(any combination of $k_1$ and $k_2$) contained in the factor.
   By explicit examination of the available scalar variables, their crossing
properties and their behavior under charge conjugation we see that we can
write the factors of the coefficients in the general form
\begin{eqnarray}\label{coefexp}
&&A_0 = a_0,\,\, A_1 = 0,\,\,
A_2 = a_{2a}k_1 \cdot k_2 +a_{2b}(k_1^2+k_2^2) +a_{2c}[(k_1+k_2) \cdot P]^2,
\,\, A_3 = 0,\nonumber \\
&&C_0 = C_1 = C_2 = 0,\,\,
C_3 = c_3(k_1^2-k_2^2)(k_1+k_2)\cdot P,\nonumber\\
&&\tilde{C}_0 = 0,\,\,
\tilde{C}_1 = \tilde{c}_1 (k_1+k_2) \cdot P, \,\,
\tilde{C}_2 = 0,\nonumber\\
&&\tilde{C}_3 = \tilde{c}_{3a}(k_1+k_2) \cdot P k_1 \cdot k_2
+ \tilde{c}_{3b}(k_1+k_2) \cdot P (k_1^2+k_2^2)
+ \tilde{c}_{3c}[(k_1+k_2) \cdot P]^3,\nonumber\\
&&\tilde{E}_0=\tilde{E}_1=0,\,\, \tilde{E}_2=\tilde{e}_2(k_1^2-k_2^2).
\end{eqnarray}
   The other crossing-even and charge-conjugation-even coefficients
$B,E,F,G$ are expanded in a fashion similar to $A$, the crossing-even,
charge-conjugation-odd coefficient $D$ similar to $C$,
and the crossing-odd, charge-conjugation-odd coefficient $\tilde{D}$ similar
to $\tilde{C}$.
   $A_4$ and $B_4$ are needed for a consistent expansion to ${\cal O}(k^4)$ but
are eliminated in the solution of the equations.
   Once crossing is satisfied, time reversal imposes the same constraints
on the expansions of these coefficients as does charge conjugation.

   These expansions of the coefficients are substituted into
Eqs.\ (\ref{geqns}) and the equations are solved order by order in powers of
$k$.
   Thus for example the ${\cal O}(k^0)$ solution gives $a_0=b_0=0$.
   One has to be careful in the solution not to introduce kinematic
singularities in the overall factors, which dictates to some extent which
coefficients are eliminated and which are saved \cite{Bardeen,Tarrach}.
   Furthermore at some points it is necessary to use the expansions of
Eq.\ (\ref{coefexp}) to cancel kinematic factors which would otherwise end up
in the denominator \cite{Tarrach}.
   The result is
\begin{equation}
\tilde{M}_B^{\mu\nu} = {\cal O}(k^2)
+{\cal O}(k^4) + \cdots,
\end{equation}
where the terms of increasing orders of $k$, taken on shell, are the following:

${\cal O}(k^2)$:
\begin{eqnarray}
\label{ok2}
&+& g_0[k_2^\mu k_1^\nu - k_1 \cdot k_2 g^{\mu\nu}] \nonumber \\
&+& \tilde{c}_1 [ (k_1+k_2) \cdot P (P^\mu k_1^\nu + k_2^\mu P^\nu)
-2k_1 \cdot k_2P^\mu P^\nu -2 k_1 \cdot P k_2 \cdot P g^{\mu\nu}],
\end{eqnarray}

${\cal O}(k^4)$:
\begin{eqnarray}
\label{ok4}
&+& [g_{2a}k_1 \cdot k_2 + g_{2b} (k_1^2 + k_2^2) + 4g_{2c}
k_1 \cdot P k_2 \cdot P][k_2^\mu k_1^\nu - k_1 \cdot k_2 g^{\mu\nu}]
\nonumber \\
&+& [ \tilde{c}_{3a}k_1 \cdot k_2 + \tilde{c}_{3b} (k_1^2 + k_2^2) +
4\tilde{c}_{3c} k_1 \cdot P k_2 \cdot P] \nonumber \\
&\quad& \times [ (k_1+k_2) \cdot P
(P^\mu k_1^\nu + k_2^\mu P^\nu)
-2k_1 \cdot k_2P^\mu P^\nu -2k_1 \cdot P k_2 \cdot P g^{\mu\nu} ]
\nonumber \\
&+& 2\tilde{e}_2 [k_1^2k_2^2g^{\mu\nu}+k_1 \cdot k_2k_1^\mu k_2^\nu - k_2^2
k_1^\mu k_1^\nu - k_1^2k_2^\mu k_2^\nu] \nonumber \\
&+& c_3 [ (k_1^2+k_2^2)(2 k_1 \cdot P k_2 \cdot P g^{\mu\nu}
- 2 k_1 \cdot k_2 P^\mu P^\nu ) + (k_1+k_2) \cdot P (k_1^2-k_2^2)
(P^\mu k_1^\nu - k_2^\mu P^\nu) \nonumber \\
&\quad& + 2(k_1+k_2) \cdot P k_1 \cdot k_2 (P^\mu k_2^\nu + k_1^\mu P^\nu)
-4 k_1 \cdot P k_2 \cdot P (k_1^\mu k_1^\nu + k_2^\mu k_2^\nu) ]
\nonumber \\
&+& 2 d_3 [-2k_1^2k_2^2 P^\mu P^\nu -2 k_1 \cdot P k_2 \cdot P k_1^\mu k_2^\nu
+ k_1^2 (k_1+k_2) \cdot P P^\mu k_2^\nu
+ k_2^2 (k_1+k_2) \cdot P k_1^\mu P^\nu ].
\end{eqnarray}

   Thus we see that there are in general two structure-dependent constants at
${\cal O}(k^2)$ which are related to the electromagnetic polarizabilities
$\bar{\alpha}$ and $\bar{\beta}$ encountered in real Compton scattering by
\begin{equation}
\label{polarizabilities}
\bar{\alpha}=-\frac{e^2}{8\pi m}(g_0+8m^2\tilde{c}_1),\quad
\bar{\beta}=\frac{e^2g_0}{8\pi m}.
\end{equation}
   At ${\cal O}(k^4)$ we find nine structure-dependent constants of which
four, $g_{2a}, g_{2c},\tilde{c}_{3a}, \tilde{c}_{3c}$  can, in principle,
be obtained in real Compton scattering.
   The determination of the other five constants would require virtual photons
as a probe.

   We should re-emphasize that while the general form of
$\tilde{M}_B^{\mu\nu}$ is the same in each case the numerical values of the
structure-dependent constants are defined only with respect to the particular
choice of $M_A^{\mu\nu} + M_B^{\mu\nu}$ and thus will be different for the
three cases considered here, SPA, WT1, WT2, or for any other.

   The fact that there are no structure-dependent terms involving odd powers
of $k$ is due to the symmetry with respect to charge conjugation or time
reversal.
   Relaxing this constraint would, for example, lead to one additional
structure at ${\cal O}(k^3)$ of the form:

${\cal O}(k^3)$ (Charge conjugation or time reversal violating):
\begin{eqnarray}
&+& \tilde{e}_1 \lbrace (k_1+k_2) \cdot P [k_1^\mu k_1^\nu - k_2^\mu k_2^\nu -
(k_1^2-k_2^2) g^{\mu\nu}] \nonumber \\
&\quad& +2k_1 \cdot k_2 (P^\mu k_2^\nu -k_1^\mu P^\nu) - 2k_2^2P^\mu k_1^\nu
+2k_1^2k_2^\mu P^\nu \rbrace.
\end{eqnarray}
Such a term could be seen only with virtual photons, but in an experiment
sufficiently sensitive to see the ${\cal O}(k^4)$ terms, it conceivably might
be possible to set limits on such charge conjugation or time reversal
violating pieces.

   It is also interesting to observe that there are no terms of
${\cal O}(k)$, so in this case the combination of gauge invariance and
crossing symmetry actually determines the amplitude to one higher order in
$k$ than in the usual soft-photon case.
   This implies that, for example, differences between two possibilities
will here first appear at ${\cal O}(k^2)$ rather than the usual ${\cal O}(k)$.
   We emphasize that this observation is valid independent of whether
charge conjugation is a symmetry of the underlying interaction or not.

   Let us finally comment on the off-shell case.
   We now have two additional scalar variables, namely, $(k_1-k_2)\cdot P =
p_f^2-p_i^2$ and $P^2$, or equivalently, $p_i^2+p_f^2$.
   If we restrict ourselves only to order ${\cal O}(k^2)$, $C_1$ and
$D_1$ pick up additional terms proportional to $(k_1-k_2)\cdot P$
whereas $A_2$ and $B_2$ obtain extra terms proportional to
$[(k_1-k_2)\cdot P]^2$.
   The result, which replaces Eq.\ (\ref{ok2}), is found to be:

${\cal O}(k^2)$ (Off shell):
\begin{eqnarray}
\label{ok2os}
&+& g_0[k_2^\mu k_1^\nu - k_1 \cdot k_2 g^{\mu\nu}] \nonumber \\
&+& \tilde{c}_1 [ (k_1+k_2) \cdot P (P^\mu k_1^\nu + k_2^\mu P^\nu)
-2k_1 \cdot k_2 P^\mu P^\nu -\frac{1}{2}[(k_1+k_2) \cdot P]^2 g^{\mu\nu}]
\nonumber \\
&+& \tilde{c}_1 (k_1-k_2) \cdot P [\frac{1}{2} (k_1-k_2) \cdot P g^{\mu\nu}
-P^\mu k_1^\nu +k_2^\mu P^\nu].
\end{eqnarray}
   This result differs from the original one in that the last term in the
second line is written differently, and in the third line which is new.
   The two coefficients are still functions of $p_i^2+p_f^2$ and reduce to
the on-shell values of Eq.\ (\ref{ok2}) for $p_i^2+p_f^2=2m^2$.

   The above serves as a demonstration that the off-shell case can, in
principle, be treated in a completely analogous fashion.
   When trying to solve the off-shell case it is important to realize that
$p_f^2-p_i^2$ is actually of ${\cal O}(k)$.
   However, since the amplitude involving off-shell pions does not correspond
to a physically observable situation, we do not extend the analysis to higher
orders.

\section{Summary}
\label{sum}

   We have considered various possible approaches to deriving low-energy
expressions for virtual Compton scattering off a charged spin-zero target,
in particular the soft-photon approach and the Ward-Takahashi approach.
   The conditions imposed on the full on-shell amplitude by gauge invariance,
crossing symmetry, and the discrete symmetries are exactly the same in the
two approaches so the physical content and the result for the full amplitude
will be exactly the same.

   However, the natural way to separate the amplitude into pieces which
contain all the singularities as $k_1 \rightarrow 0$ or $k_2 \rightarrow 0$
is different in the various cases.
   These different choices in effect amount to moving pieces which are
non-singular, separately gauge invariant, and at least of ${\cal O}(k)$,
back and forth between the part containing the singularities and the rest.

   We have also identified the possible structure-dependent constants allowed
by gauge invariance, the discrete symmetries, and crossing through
${\cal O}(k^4)$.
   The general form obtained is independent of the choice of approach for
separating out the singular terms but the numerical values of the
coefficients will depend on this choice.
   This means that such structure-dependent constants, e.g., the
electromagnetic polarizabilities, must be defined with respect to a
particular choice of the pole terms, and the numerical values will depend
on this choice.
   At ${\cal O}(k^2)$ we found two structures which can already be seen
in real Compton scattering and which are related to the usual electric
and magnetic polarizabilities.
   Due to symmetry with respect to charge conjugation there are no
terms containing odd powers of $k$.
   At ${\cal O}(k^4)$ we found nine new structure-dependent terms.
   Four of these terms can, in principle, be determined in real Compton
scattering, whereas the other five terms require virtual photons.

   Finally, we saw that in this case gauge invariance and crossing symmetry
imply that there are no terms of ${\cal O}(k)$, independent of whether
charge conjugation is a good symmetry or not, so the low-energy expression
is good to one higher order than is usual in radiative processes.

\acknowledgements
   H.\ W.\ Fearing would like to thank SFB 201 of the Deutsche
Forschungsgemeinschaft for its hospitality and financial support during
his stay at Mainz.
   S.\ Scherer thanks the TRIUMF theory group for the kind hospitality
and financial support.
   The collaboration between TRIUMF and the Institute for Nuclear Physics at
Mainz is supported in part by a grant from NATO.
   This work was also supported in part by the Deutsche Forschungsgemeinschaft
(S.S.) and the Natural Sciences and Engineering Research Council of
Canada (H.W.F.).

\end{document}